\documentclass[nocompress]{spie}  
\usepackage[]{graphicx}
\usepackage{amssymb}
\usepackage{amsmath}

\title{Developing a second generation Laue lens prototype: high reflectivity crystals and accurate assembly} 

\author{Nicolas M. Barri{\`e}re\supit{a}, John A. Tomsick\supit{a}, Steven E. Boggs\supit{a}, Alexander Lowell\supit{a},
and Peter von Ballmoos\supit{b}
\skiplinehalf
\supit{a}Space Sciences Laboratory, 7 Gauss Way, University of California, Berkeley, CA 94720-7450 - USA \\
\supit{b}Institut de Recherche en Astrophysique et Planetology, UMR 5277,  9 av. du Colonel Roche, 31028 Toulouse - France
}

\authorinfo{Corresponding author: N.B. \\E-mail: barriere {\it at}  ssl.berkeley.edu}

\begin{document} 
  \maketitle 

\begin{abstract}
Laue lenses are an emerging technology that will enhance gamma-ray telescope sensitivity by one to two orders of magnitude in selected energy bands of the $\sim$ 100 keV to $\sim$ 1.5 MeV range. This optic would be particularly well adapted to the observation of faint gamma ray lines, as required for the study of Supernovae and Galactic positron annihilation. It could also prove very useful for the study of hard X-ray tails from a variety of compact objects, especially making a difference by providing sufficient sensitivity for polarization to be measured by the focal plane detector. Our group has been addressing the two key issues relevant to improve performance with respect to the first generation of Laue lens prototypes: obtaining large numbers of efficient crystals and developing a method to fix them with accurate orientation and dense packing factor onto a substrate. We present preliminary results of an on-going study aiming to enable a large number of crystals suitable for diffraction at energies above 500 keV. In addition, we show the first results of the Laue lens prototype assembled using our beamline at SSL/UC Berkeley, which demonstrates our ability to orient and glue crystals with accuracy of a few arcsec, as required for an efficient Laue lens telescope.
\end{abstract}


\keywords{Telescope, Soft gamma rays, Focusing optics, Crystals, Technological development}

\section{INTRODUCTION}
\label{sec:intro} 
A Laue lens is a single reflexion concentrator based on Bragg diffraction in the volume of a large number of small crystal slabs oriented in order to diffract incident radiation towards a common point [\citenum{lund.1992ft}] . Although the Bragg diffraction is very chromatic, the use of either mosaic crystals or crystals having curved diffraction planes allows Laue lenses to cover broad bandpasses, of the order of a few hundreds of keV (as shown for instance in Ref.~[\citenum{barriere.2010fk}] and [\citenum{frontera.2005uq}]) within the range from $\sim$ 100 keV to $\sim$ 1.5 MeV. 
A Laue lens is a single reflexion concentrator but it is possible to reconstruct an image of the field of view with an angular resolution of the order of  an arcmin [\citenum{barriere.2009fk}], depending on the crystals size, their orientation accuracy and their mosaicity\footnote{The {\it mosaicity} refers to the full width at half maximum of the angular distribution of crystallite, the tiny perfect crystals that compose a mosaic crystal according to Darwin's model. By extension this term is used to describe the angular spread of any crystal, mosaic crystals or crystals having curved diffracting planes.}. Another feature of Laue lenses is that they do not change the polarization of the incident signal, which can then be analyzed in the focal plane instrument [\citenum{barriere.2008}].
Laue lens telescopes are expected to provide an increase in sensitivity by one to two orders of magnitude with respect to existing instruments due to the fact that they focus from a large collecting area onto a small detector volume, hence increasing dramatically the signal-to-noise ratio.

Several fields of soft gamma-ray astronomy would benefit of the development of Laue lenses. One such topic is the study of explosion mechanism and progenitor nature of Type Ia supernovae (SNeIa). The spectroscopy and light curve of  the line at 847 keV emitted by the decay chain of $^{56}$Ni, which is massively synthesized in SNeIa, would discriminate between the currently competing models. A Laue lens telescope as featured in the DUAL mission [\citenum{von-ballmoos.2010zr, boggs.2010uq}]  (proposed  to the European Space Agency for the third medium class mission AO of the cosmic Cosmic Vision program) could reach a sensitivity of 2$\times$10$^{-6}$ ph/s/cm$^2$ (3 $\sigma$, 1 Ms) in the 3-\% broadened line at 847 keV, enough to observe a dozen events each year out to $\sim$40 Mpc and make a breakthrough in our understanding of their physics (see e.g. Ref.~[\citenum{barriere.2010rt, leising.2005qr, knodlseder.2009kx}]).

Another topic is the study of the electron-positron annihilation radiation at 511 keV. This line has been observed for more than 30 years from the Galactic center, yet it is still unclear whether known sources can account for all the 10$^{43}$ positrons that annihilate every second in the Galactic bulge [\citenum{knodlseder.2005rc}]. New observational clues are needed but both improved sensitivity and improved angular resolution are required. A Laue lens telescope could provide the way to probe small sky regions to check for structure in the emission and probe some candidate sources, like X-ray binaries. 

A Laue lens observing in a selection of narrow bands of the 100 keV to 1MeV domain could bring valuable clues concerning the emission mechanisms - yet poorly understood - of blazars and AGNs. The determination for a couple of AGNs of the turnover from synchrotron to inverse-Compton  branches would constrain the emission models. Many non-blazar AGN show an undisturbed power law up to 200 keV. From accretion theory, a turn-over in the spectrum is expected, but a precise measurement has not been possible in many cases. Sources like Cen A seem to break somewhere around 500 keV [\citenum{abdo.2010fk}], an energy range hardly accessible so far. A Laue lens telescope would give important insight into the energetics close to the central engine of super massive black holes.

Observations of the hard X-ray tails in Galactic black-hole binaries would also bring clues on the physics of these objects. At high mass accretion rates when these systems are in the ``Steep Power-Law'' state, they exhibit a power-law spectral component that extends to at least hundreds of keV [\citenum{grove.1998kx}]. No spectral break has been detected yet, due to the lack of sensitivity of present instruments above 100 keV. The production mechanism of this spectral component is currently unknown. In the``Hard'' state, a spectral break is often seen, but observations with INTEGRAL indicate the presence of at least two components in the 100-600 keV bandpass [\citenum{bouchet.2009uq}], indicating that our understanding of this part of the spectrum is still incomplete.

In order to meet the sensitivity requirements for these objectives, the best possible efficiency is needed for the Laue lens. That translates into crystals with high reflectivity, densely packed, and accurately oriented. We have taken up the challenge of developing such lens at SSL. We report in this paper on the status of this effort. In section \ref{sec:crystals}, we present the work we are conducting to enable efficient crystals for high energy Laue lenses. Section \ref{sec:LaueLens} presents the crystal orientation requirements to build scientifically viable Laue lenses. Then, in section \ref{sec:assembly}, we present the Laue lens assembly facility that we have constructed at SSL, and finally we show in section \ref{sec:results} the first results we obtained for gluing crystals.

\section{CRYSTAL STUDY} 
\label{sec:crystals} 

\begin{figure}[t]
\begin{center}
\includegraphics[width=0.65\textwidth]{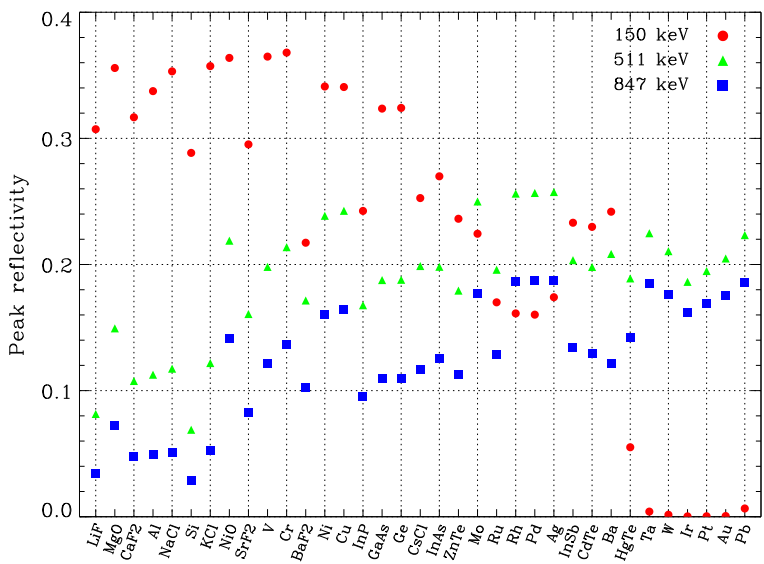}
\caption{Reflectivity at 150, 511 and 847 keV (respectively red, green and blue points) of a selection of crystals that are potentially interesting for the realization of a Laue lens. The crystals are sorted from left to right by increasing mean atomic number. The reflectivity was calculated assuming they are mosaic crystals with 45 arcsec of mosaicity. The dynamical theory of diffraction was used with different crystallite size for each energy [\citenum{Barriere:he5432, barriere_2010b}]: 30 $\mu$m, 60 $\mu$m and 90 $\mu$m respectively for 150 keV, 511 keV and 847 keV. The thickness is computed to maximize the peak reflectivity within the limits 2 mm $\leqslant$ T$_0$ $\leqslant$ 15 mm.}
\label{fig:reflectivity}
\end{center}
\end{figure}
\begin{figure}[t]
\begin{center}
\includegraphics[width=0.45\textwidth]{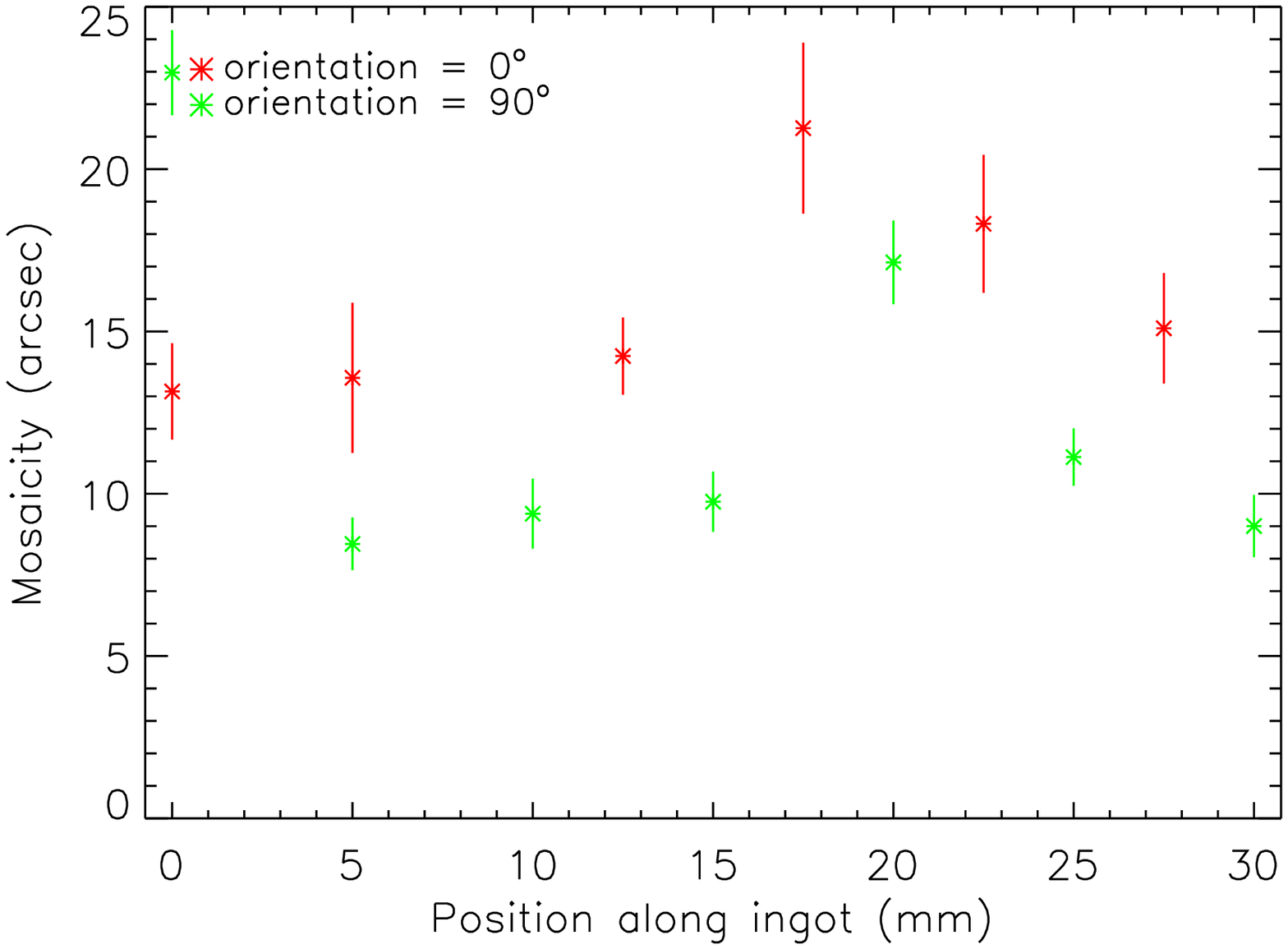}
\includegraphics[width=0.45\textwidth]{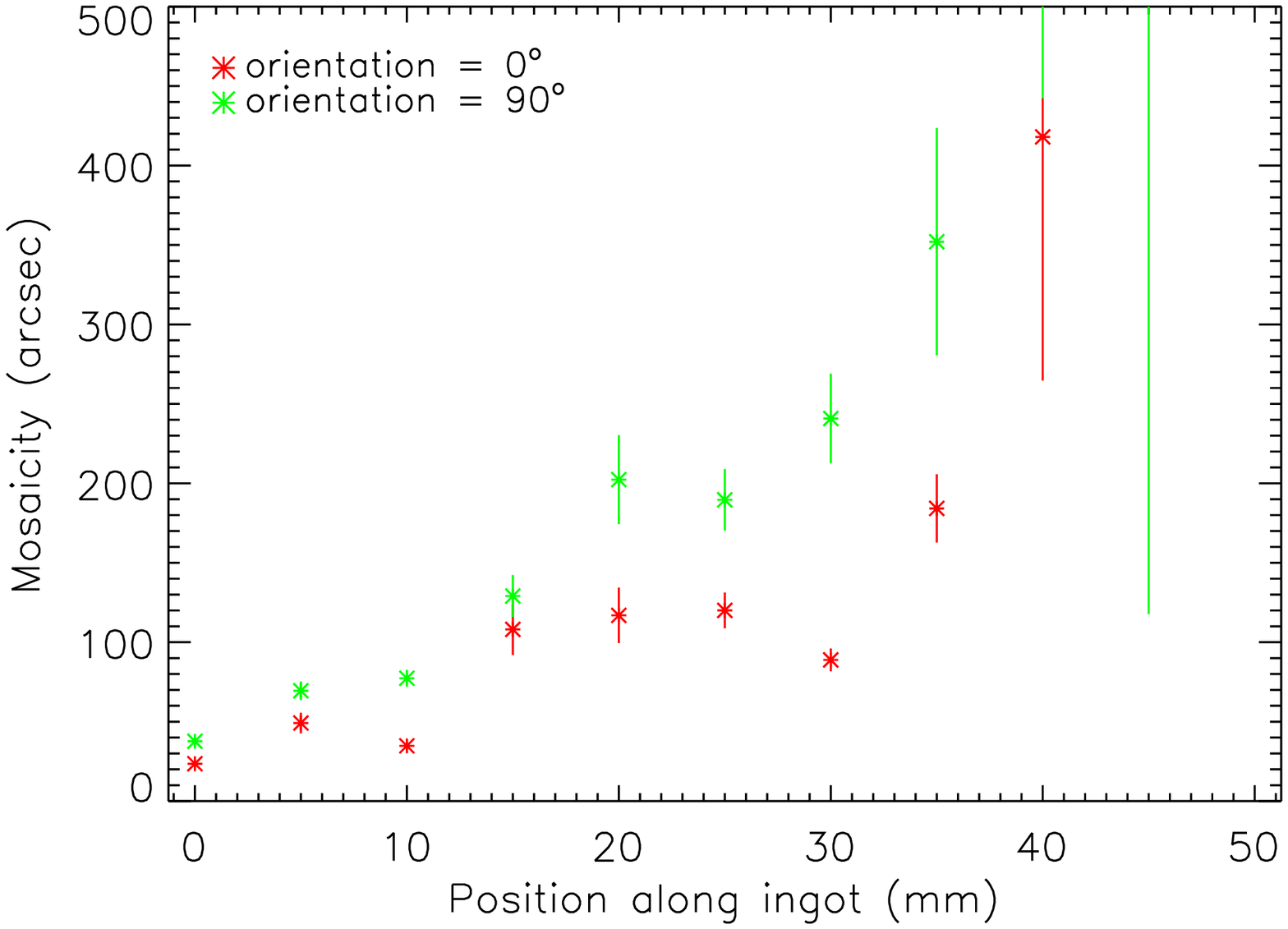}
\includegraphics[width=0.45\textwidth]{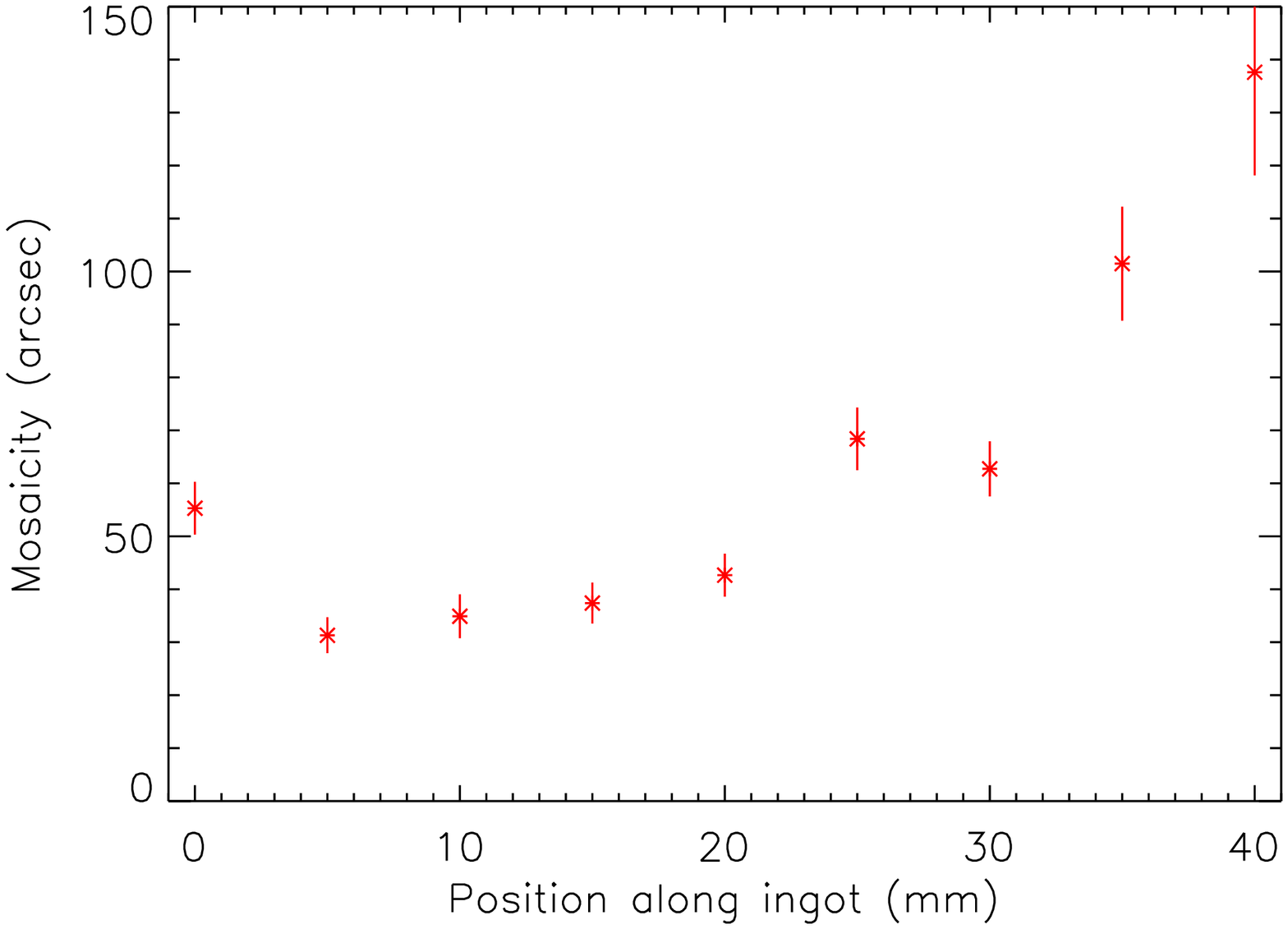}
\caption{Mosaicity versus position for three ingots produced by Mateck and measured in a 517 keV beam at ILL / DIGRA. On the left results are shown for a silver ingot measured across the diameter (10.7 mm) using the refelxion 200. This crystal was measured twice, rotating it of 90$^{\circ}$ about its axis between the two series of measurements. On the right, results are shown for a Rh 111 ingot (in two directions as well), measured across it diameter of 11 mm. The bottom plot shows results for a Pb 200 crystal, measured across its diameter of 12 mm.}
\label{fig:mosaicityingots}
\end{center}
\end{figure}

For several years, we have been working to identify crystals suitable to build {\it efficient} Laue lenses. For this purpose the first task was to establish a list of materials fulfilling basic requirements: the material must exist in a crystalline state at room temperature, it must not be ductile, and it must not be (too) toxic or radioactive.  Then, we excluded from the list all the crystal lattices not based on a cube as they are far more difficult to work with or show lower diffraction efficiency. Crystals composed of different atoms tend to have larger lattice parameters than pure materials, which dramatically decreases their diffraction efficiency. So, we limited our search to pure materials and bi-component crystals. Figure \ref{fig:reflectivity} is an updated version of the plot published in Ref.~[\citenum{Barriere:he5432}]. It shows the peak reflectivity of the crystals currently in our list for three different energies. More two-component crystals might join this list, but it is complete regarding pure materials. 

Once this theoretical list was established, one had to determine wether these materials are available in single crystals with large volume, and, if so, with what mosaicity. Depending on the project (focal length, energy band), the optimal mosaicity ranges from 0.5 to 2 arcmin. So, over the past few years we tried to procure samples of these crystals and go to high energy X-ray or gamma ray facilities to determine their quality. So far, we have been able to measure samples of: CaF$_2$, Si, Cu, InP, GaAs, Ge, Rh, Ag, CdTe, Ta, W, Ir, Pt, Au and Pb.
Amongst this list, only Ta, W, Ir and Pt were not of sufficient quality (i.e. mosaicity larger than 10 arcmin and very irregular), although for most of them, only one sample was measured. These crystals are of interest for energies above $\sim$ 750 keV thanks to their high electron density. However, they are very difficult to grow properly due to their high melting point, and, as a consequence, are very expensive. The advent of one of these four crystals for use in Laue lenses would further improve the overall lens efficiency, as having several crystals that diffract the same energy from a given radius allows the most efficient one to be chosen.

Recently, we have focused our interest to high energy Laue lenses working around 511 keV and 847 keV [\citenum{barriere_2010b}] . These lenses are composed of Rh, Ag, Pb, Cu and Ge with an option to replace Ag by Au in the 847 keV lens (these two crystals have very similar lattice parameters, and thus they can be swapped without any design change). Since Cu and Ge have already been proven to be available for use in Laue lenses [\citenum{rousselle.2011fk, von-ballmoos.2004sf}], we are focusing on enabling Ag, Rh and Pb.  In all three cases, several cut samples or raw ingots were already measured with mosaicity ranging between 0.5 and 1 arcmin. We now want to establish the time, cost, and overall difficulty of producing cut pieces with given orientation, thickness, and mosaicity. We ordered 5 pieces of Rh111, Rh200, Rh220, Rh311, Ag111, Ag200, Ag220, Pb111, Pb200 and  Pb220 (a total of 50 crystals) of 5$\times$5 mm$^2$ and 3 to 5 mm thickness from Mateck (Juelich, Germany). The aim is to obtain crystals with average mosaicity lower than 1.5 arcmin, so the rejection threshold for raw ingots is set to 1 arcmin (the cut is expected to induce damages). To reach this goal, we are engaged in an iterative process between ingots production and characterization in order to refine the ingot growth technique. We also check the mosaicity of the cut pieces as they are produced in order for Mateck to refine the cut parameters if necessary.

Figure \ref{fig:mosaicityingots} shows the mosaicity as function of position along the ingot axis for three samples recently measured at the Institute Laue Langevin (Grenoble, France) using the DIGRA beamline specifically developed for bulk characterization of crystals. The ingots were probed using a 5$\times$7 mm beam tuned at 517 keV in order to average the mosaicity over a large volume. As we can see, these three ingots fulfill our requirement of having a mosaicity lower than 1 arcmin (=60 arcsec), at least over a part of their length. We transmitted this information to Mateck where the ingots will be cut. Cut crystals will be extracted out of the volumes having a mosaicity lower than the 1 arcmin threshold. The knowledge of the bulk mosaicity in the ingot before the crystals are cut is precious to determine the best set of cut parameters.

\section{CRYSTAL ORIENTATION TOLERANCES}
\label{sec:LaueLens} 

Let's first define the referential for the orientation of a crystal on the lens substrate. $\Omega_B$, $\Omega_R$ and $\Omega_Z$ are the three axes of rotation defined as:  \\
$\bullet$ $\Omega_B$: in the lens plane, perpendicular to the radius connecting the crystal center to the lens center. This is the Bragg angle. \\
$\bullet$ $\Omega_R$: in the lens plane,  parallel to the radius connecting the crystal center to the lens center. \\
$\bullet$ $\Omega_Z$ perpendicular to the lens plane (i.e. parallel to the optical axis of the lens). 

\begin{figure}[t]
\begin{center}
\hspace{0.3cm}
\includegraphics[width=0.335\textwidth]{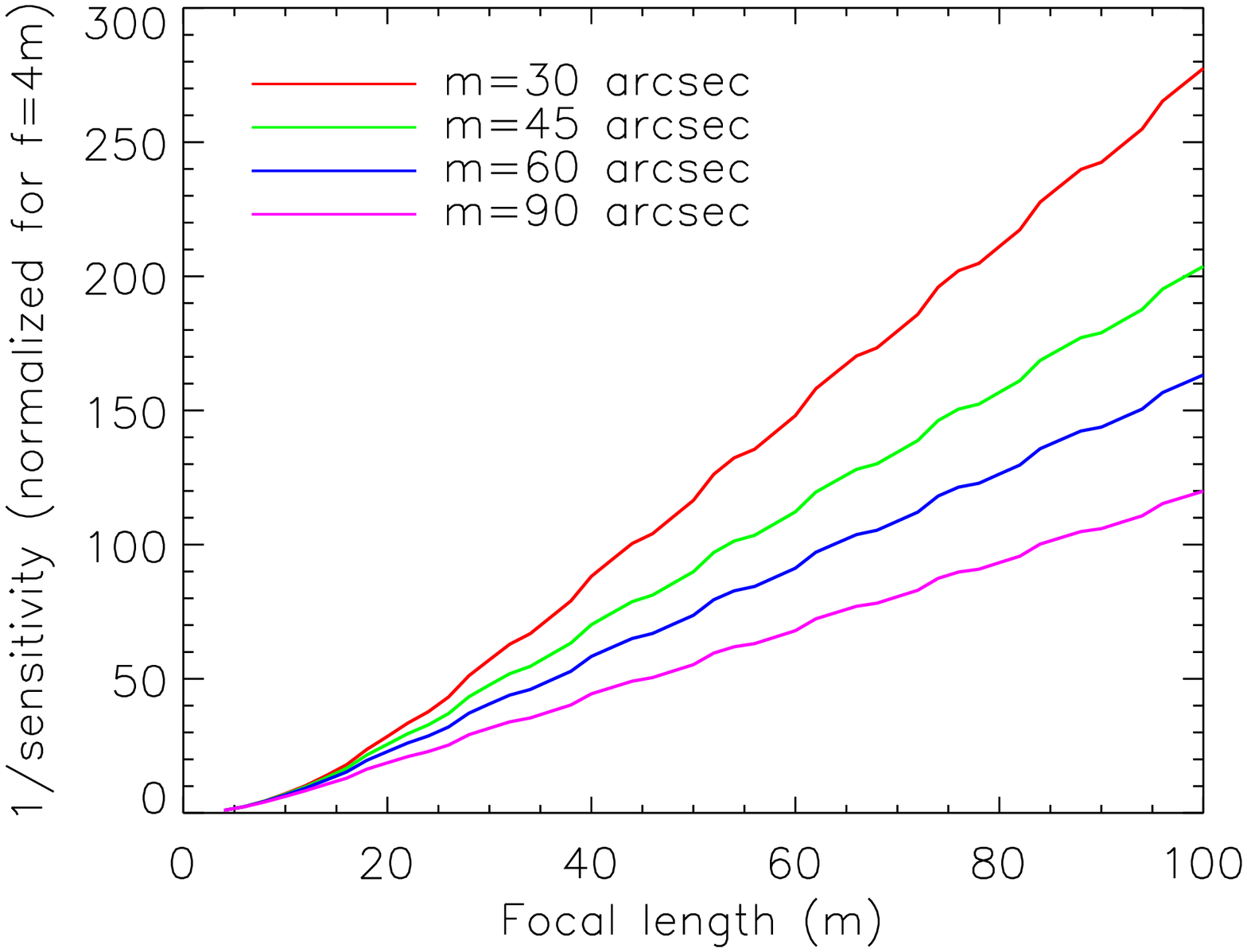}
\includegraphics[width=0.335\textwidth]{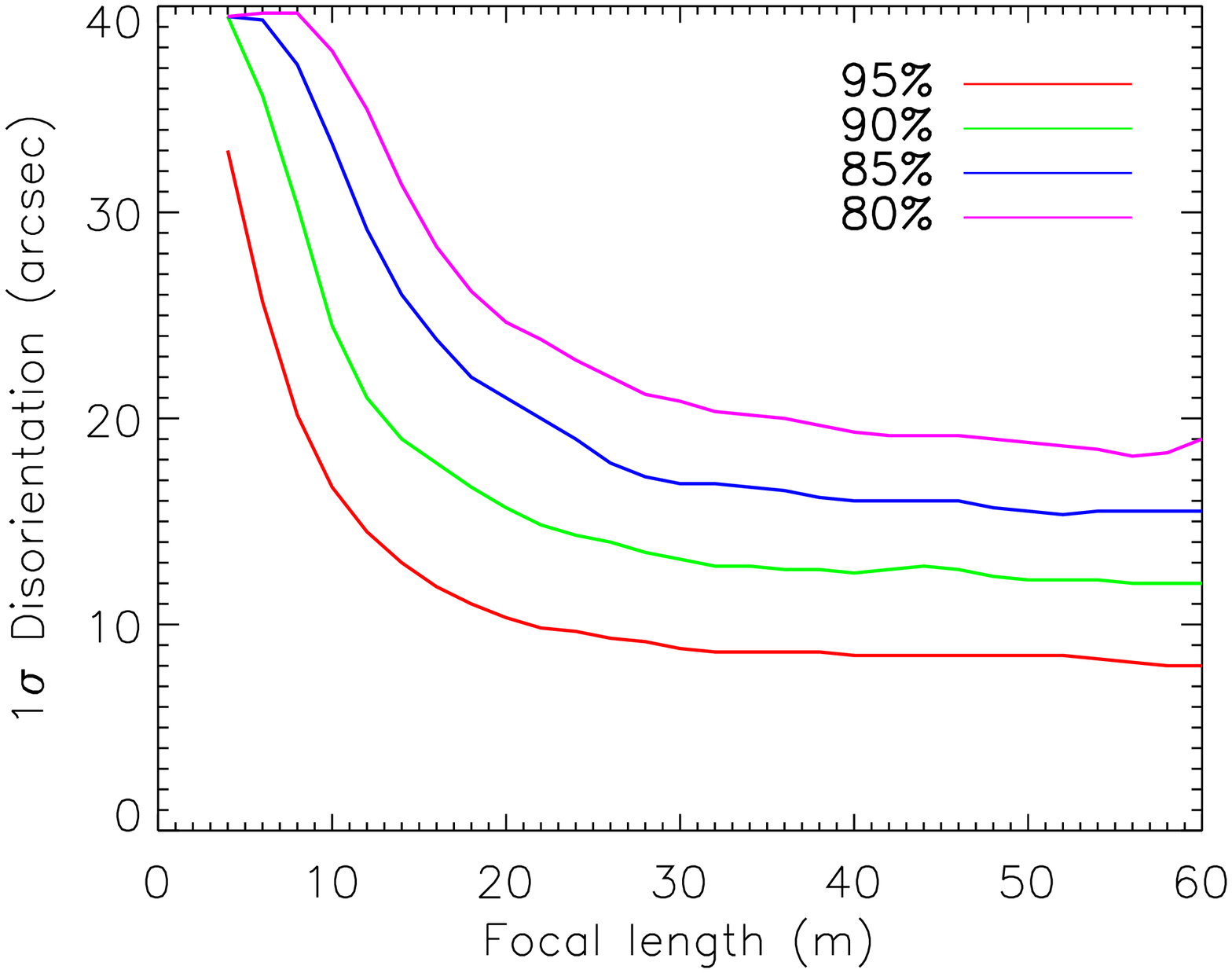}
\includegraphics[width=0.27\textwidth]{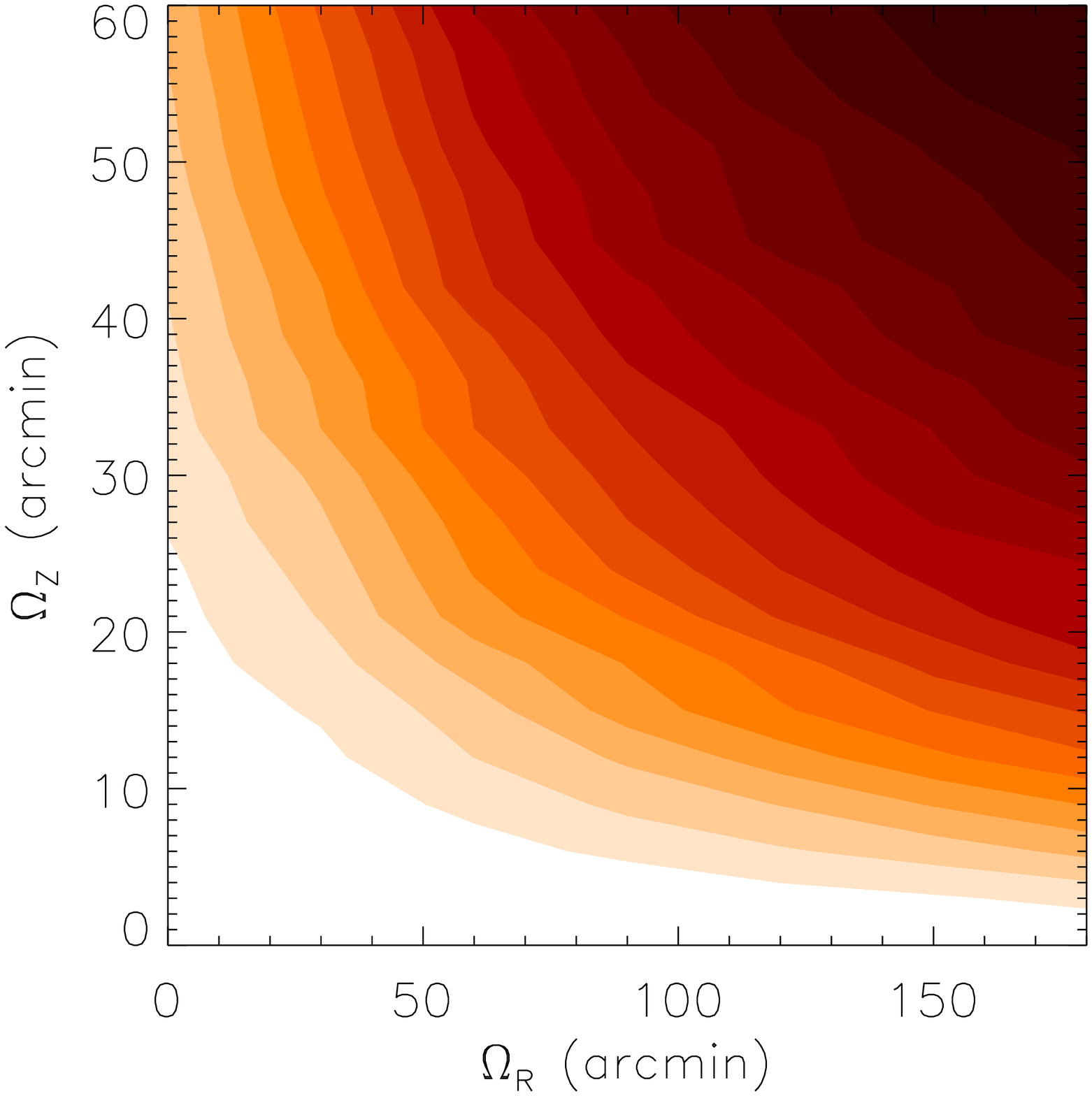}
\vspace{0.5cm}
\caption{
{\it Left:} Sensitivity of a Laue lens telescope focusing in a given energy band (700 keV - 900 keV) as a function of its focal length. The crystal size is kept constant at 10$\times$10 mm$^2$, results are shown for four different crystal mosaicities: 30, 45, 60 and 90 arcsec.
{\it Middle:} Crystal disorientation (1$\sigma$, assuming a Gaussian distribution) that produces a sensitivity loss of 5, 10, 15 and 20\% as function of the focal length. The calculation is done for a lens focusing in the 700 keV - 900 keV band, made of crystals of 10$\times$10 mm$^2$ and 45 arcsec of mosaicity. 
{\it Right:} Sensitivity loss as function of the disorientation of the two angles $\Omega_R$ and $\Omega_Z$ assuming a disorientation of 10 arcsec for $\Omega_B$. The axes show the 1 $\sigma$ values, assuming a Gaussian distribution. Each contour represents a sensitivity degradation of 4\%, the white area being the bin 0\% - 4\%.
}
\label{fig:orientationccuracy}
\end{center}
\end{figure}

Depending on the focal length and the mosaicity of the crystals employed, the requirement on crystal orientation accuracy can vary dramatically. As shown on the left panel of Figure \ref{fig:orientationccuracy}, the sensitivity of a Laue lens telescope improves with its focal length. The improvement can vary slightly depending on whether the crystal size and mosaicity are kept constant or not, but, in any case, a longer focal length produces a more sensitive telescope. One should notice, however, that a longer focal length means a larger number of crystals, and a bigger and heavier lens. The second point shown in this plot is that a smaller crystal mosaicity produces a more sensitive telescope. Again, there are subtleties (especially for focal length smaller than 10 m), but this is a general trend. The point is that a longer focal length and a smaller mosaicity call for very well oriented crystals, otherwise the performance degradation can be dramatic. 

In the following part of this discussion, we consider the case of the SNeIa Laue lens presented in Ref.~[\citenum{barriere_2010b}], i.e. with a focal length of 30 m and made of 10$\times$10 mm$^2$ crystals having a mosaicity of 45 arcsec. The middle panel of Figure \ref{fig:orientationccuracy} shows the angular deviation of the Bragg angle ($\Omega_B$), which leads to sensitivity loss of 5, 10, 15 and 20\%, as a function of the focal length. This assumes a Gaussian distribution of the angular deviation. For a 30-m focal length lens, a standard deviation of 10 arcsec already degrades the sensitivity by 5\%.

Keeping this value of 10 arcsec for $\Omega_B$, we estimate now the tolerance for the other angles. The right panel of Figure \ref{fig:orientationccuracy} shows that the requirements on  $\Omega_R$ and  $\Omega_Z$ are somewhat more relaxed. The sensitivity degradation is lower than 4\% if the angular deviations are within 15 arcmin for both $\Omega_R$ and  $\Omega_Z$.
This is not too constraining as crystals can be cut quite easily with orientation accuracy of the order of 10 arcmin. It means that we can rely on the external faces of the crystals to orient $\Omega_R$ and  $\Omega_Z$ properly.

\section{ASSEMBLY METHOD DEVELOPED AT SSL} \label{sec:sections}
\label{sec:assembly}

\begin{figure}[p]
\begin{center}
\includegraphics[width=0.9\textwidth]{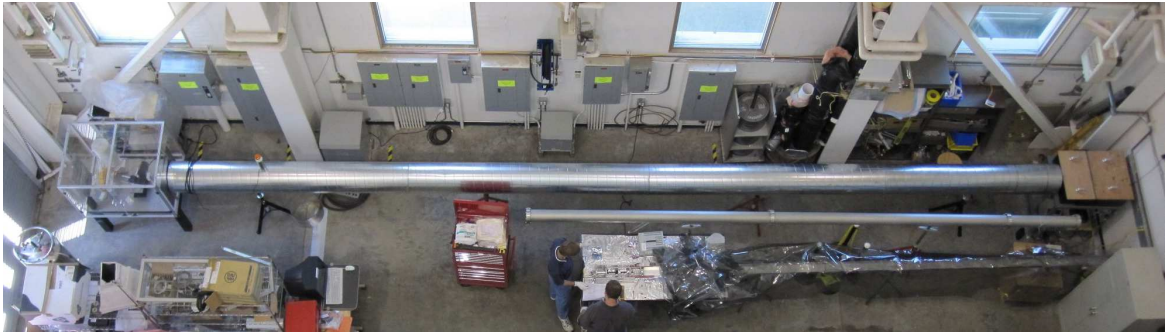}
\caption{Bird's eye view of the X-ray beamline in the high bay at SSL. The X-ray generator is enclosed in a lead casing on the right hand side of the room (wooden boards) and the Laue lens assembly station is in the plexiglass enclosure on the left hand side of the room.}
\label{fig:beamline}
\end{center}
\end{figure}

\begin{figure}[p]
\begin{center}
\includegraphics[height=6.4cm]{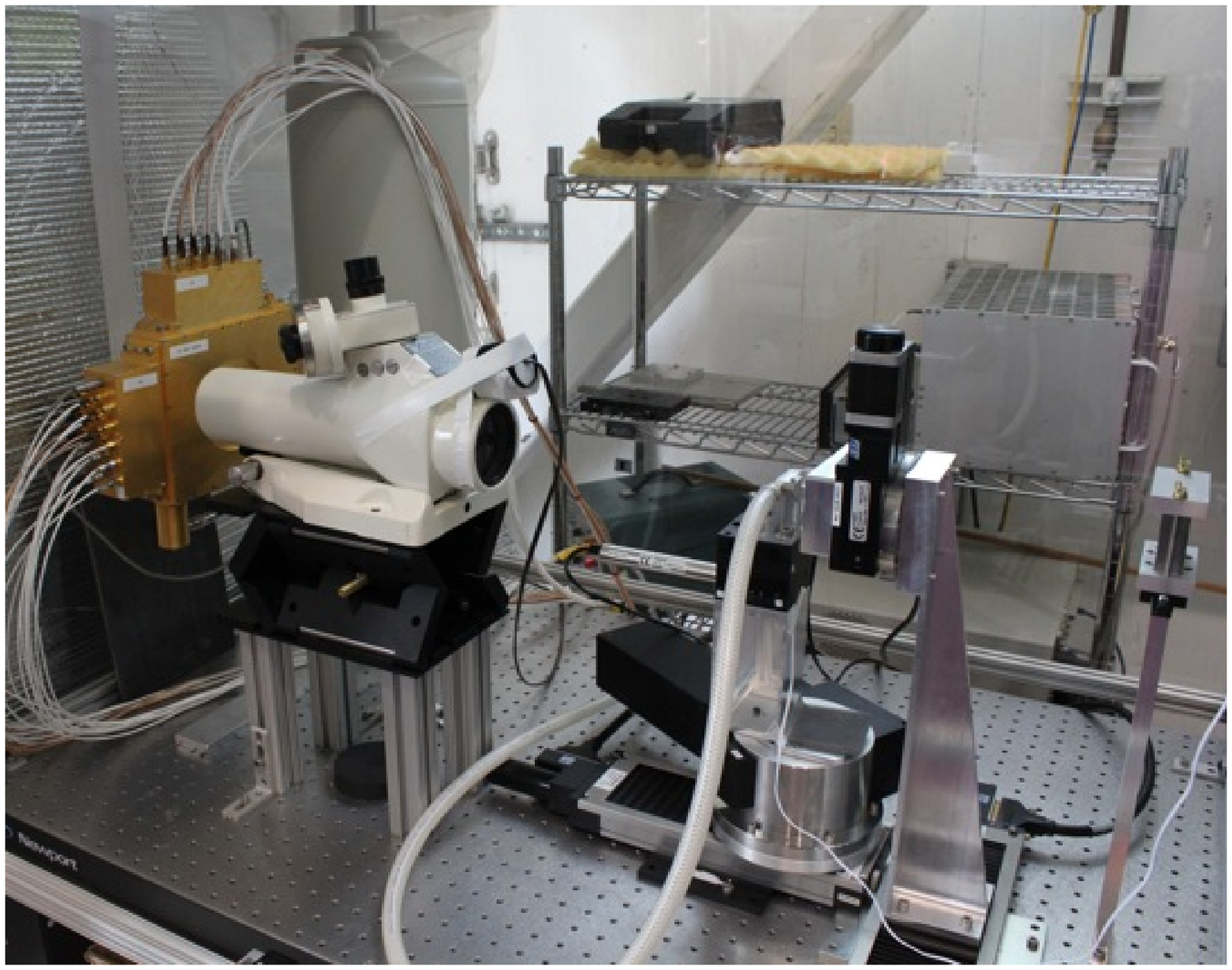}
\hspace{1cm}
\includegraphics[height=6.4cm]{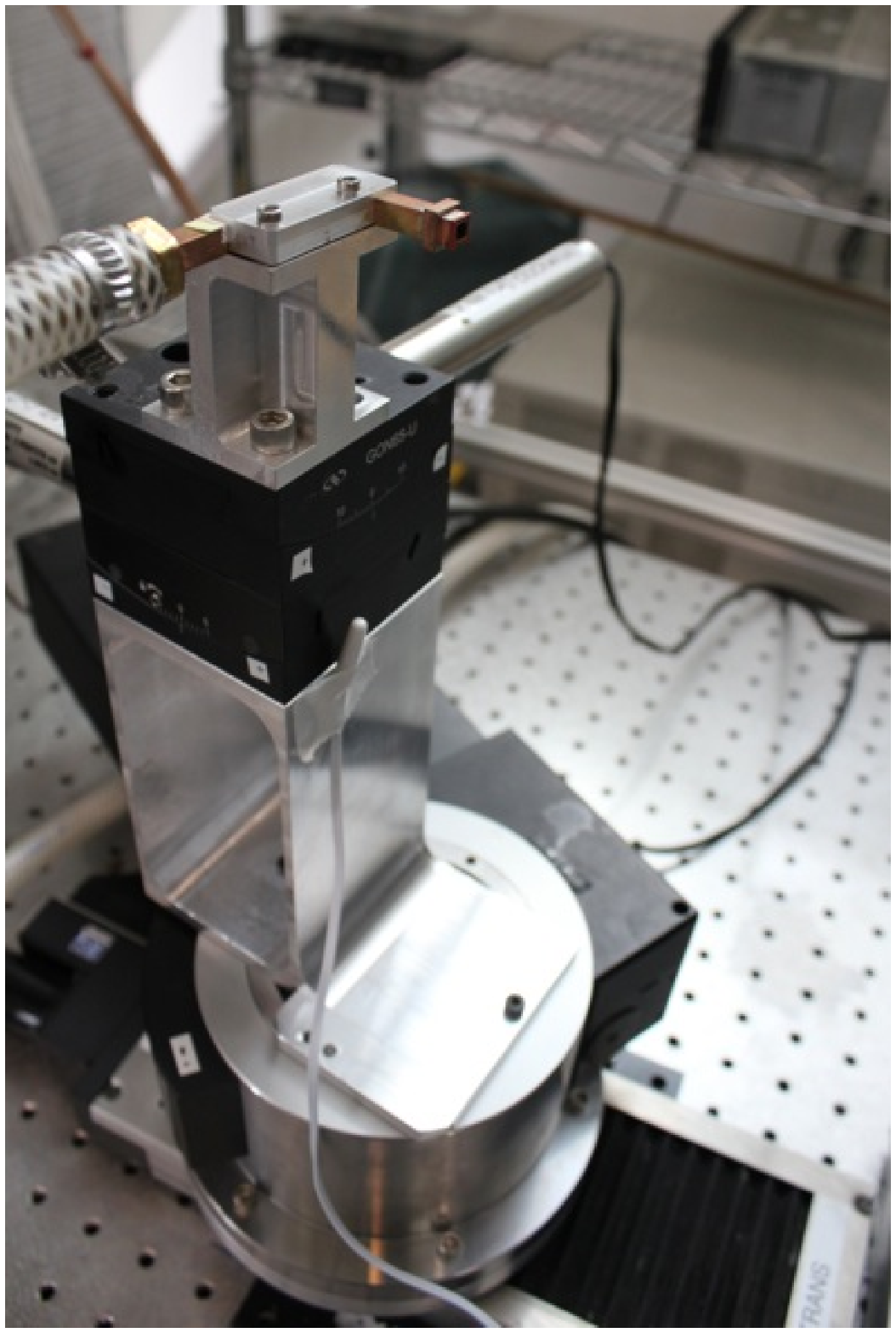}
\caption{{\it Left:} Laue lens assembly station. From the left to the right, we can see the Ge detector (golden cryostat), the autocollimator, the crystal tower, the substrate tower and the tungsten slits defining the beam size illuminating the crystal. {\it Right:} Crystal holder composed of a vacuum chuck on top of a set of rotation and translation stages.}
\label{fig:exptable}
\end{center}
\end{figure}

As briefly mentioned in Ref.~[\citenum{barriere_2010b}], a 13-m long beamline fully dedicated to the development of Laue lens was constructed at SSL (Figure \ref{fig:beamline}). It is composed at one end of an X-ray generator (XRG) working at 150 kV and 450 $\mu$A (lent by P. von Ballmoos' group, IRAP, France) and at the other end by a cross-strip high purity germanium detector. The spectrum incident on the samples is a continuum from $\sim$50 keV to 150 keV peaking at $\sim$100 keV (the low energies are attenuated by a 1.5-mm thick Cu filter). In the XRG, the electron spot on the tungsten target is $\sim$0.8 mm in diameter, which corresponds to a beam divergence of 21.5 arcsec when the slit in front of the crystal 
is set to 0.5 mm (Figure \ref{fig:exptable}). The detector, which is a prototype for the Nuclear Compton Telescope (NCT) [\citenum{bellm.2009kx}], has 19 strips on each face with a 2-mm pitch and is 11-mm thick [\citenum{coburn.2001fk}]. Software has been adapted from the NCT ground segment equipment in order to display in real time a map of the 19$\times$19 voxels, with the possibility to extract the spectrum out of a selected region and fit a Gaussian peak. A repeatability of 0.035 keV was observed in the fit of the diffracted peak of a perfect 5$\times$5$\times$3 mm$^3$ Si 111 crystal at 100 keV doing 200-s integrations, which turns into an orientation accuracy of 1.5 arcsec.

\begin{figure}[p]
\begin{center}
\includegraphics[height=6.4cm]{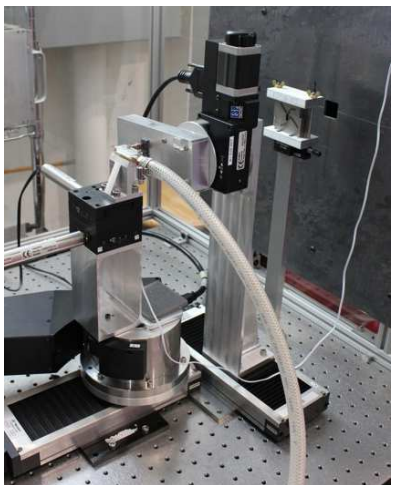}
\hspace{1cm}
\includegraphics[height=6.4cm]{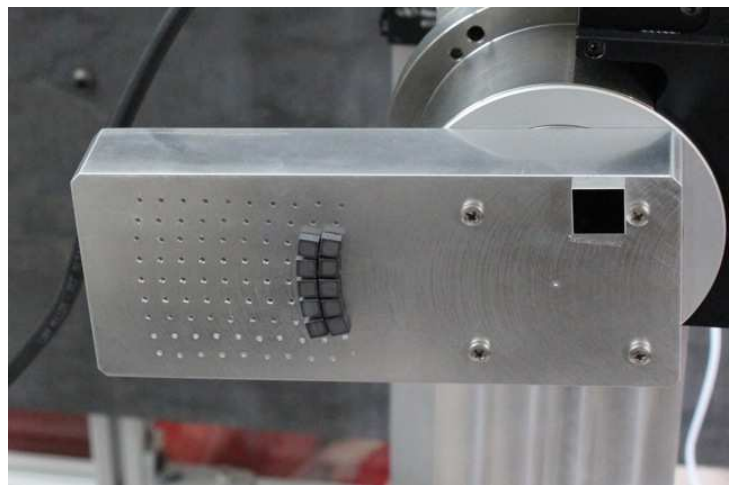}
\caption{{\it Left:} Crystal being held in position onto the substrate while the glue is curing. {\it Right:} The substrate populated with 10 Si crystals. The reference mirror viewed by the autocollimator is visible in its upper-right corner.}
\label{fig:prototype}
\end{center}
\end{figure}

The Laue lens assembly station is composed of a crystal holder (Figure \ref{fig:exptable}) and a substrate holder (Figure \ref{fig:prototype}). The crystal holder is composed of a vacuum chuck on top of a set of three rotation stages and one translation stage. It allows for the crystals to be held firmly (but temporarily) in place while they are oriented with respect to the beam and brought against the substrate for gluing. The substrate holder is composed of a translation stage (allowing for changes in the radius of the lens to be populated) and a rotation stage with an axis that defines the optical axis of the lens. This setup is actually incomplete as we are missing a rotation stage to control precisely the orientation of the substrate with respect to the beam. However, our goal was to demonstrate that the gluing of crystals can be done with high accuracy with respect to the X-ray beam, which doesn't require the substrate to be aligned with respect to the beam, as long as it is kept fixed.

The following steps describe the method we used to orient and glue crystals:
\begin{enumerate}
\item The crystal is setup at the tip of the vacuum chuck. The two angles $\Omega_R$ and $\Omega_Z$ (see section \ref{sec:LaueLens}) rely on the accuracy to which the crystals are cut and are set manually as the crystal is placed on the vacuum chuck. 
\item The crystal is pre-oriented within a few keV of the desired energy by looking at its diffracted peak energy and brought in front of the hole in the substrate where it will be glued. For this trial, we used a distance of $\sim$60 $\mu$m at the closest point.
\item The orientation of the substrate is measured with an autocollimator (Its optical axis is the reference, and consequently it is assumed fixed).
\item The crystal orientation (Bragg angle, $\Omega_B$) is then finely tuned by rocking it until the desired diffracted energy is observed on the detector. Integrations of 200 s were done to insure the aforementioned angular accuracy.
\item Some glue is injected through the substrate hole. Two-part epoxy MasterBond EP30-2 that features extremely low shrinkage upon cure of 3$\times10^{-4}$ was used for this trial.
\item The crystal is maintained in position for 6 hours before releasing slowly the vacuum and sliding back the vacuum chuck.
\item The orientation of the substrate is measured optically, and the energy diffracted by the crystal is measured. 
\end{enumerate}

 \section{RESULTS OF THE FIRST TRIAL}
 \label{sec:results}

Following the method described in section \ref{sec:assembly}, we glued 10 crystals over the course of 5 consecutive days of August 2011 (see Figure \ref{fig:prototype}). Perfect silicon 111 crystals of 5$\times$5$\times$5$\times$3 mm$^3$ produced at IKZ (Germany) were used for this trial. The substrate is a 1-inch thick aluminum plate carved from the back to leave 1 mm behind the crystals. The aim of this trial was to orient the crystals to diffract at 100 keV.

\begin{figure}[t]
\begin{center}
\includegraphics[width=0.48\textwidth]{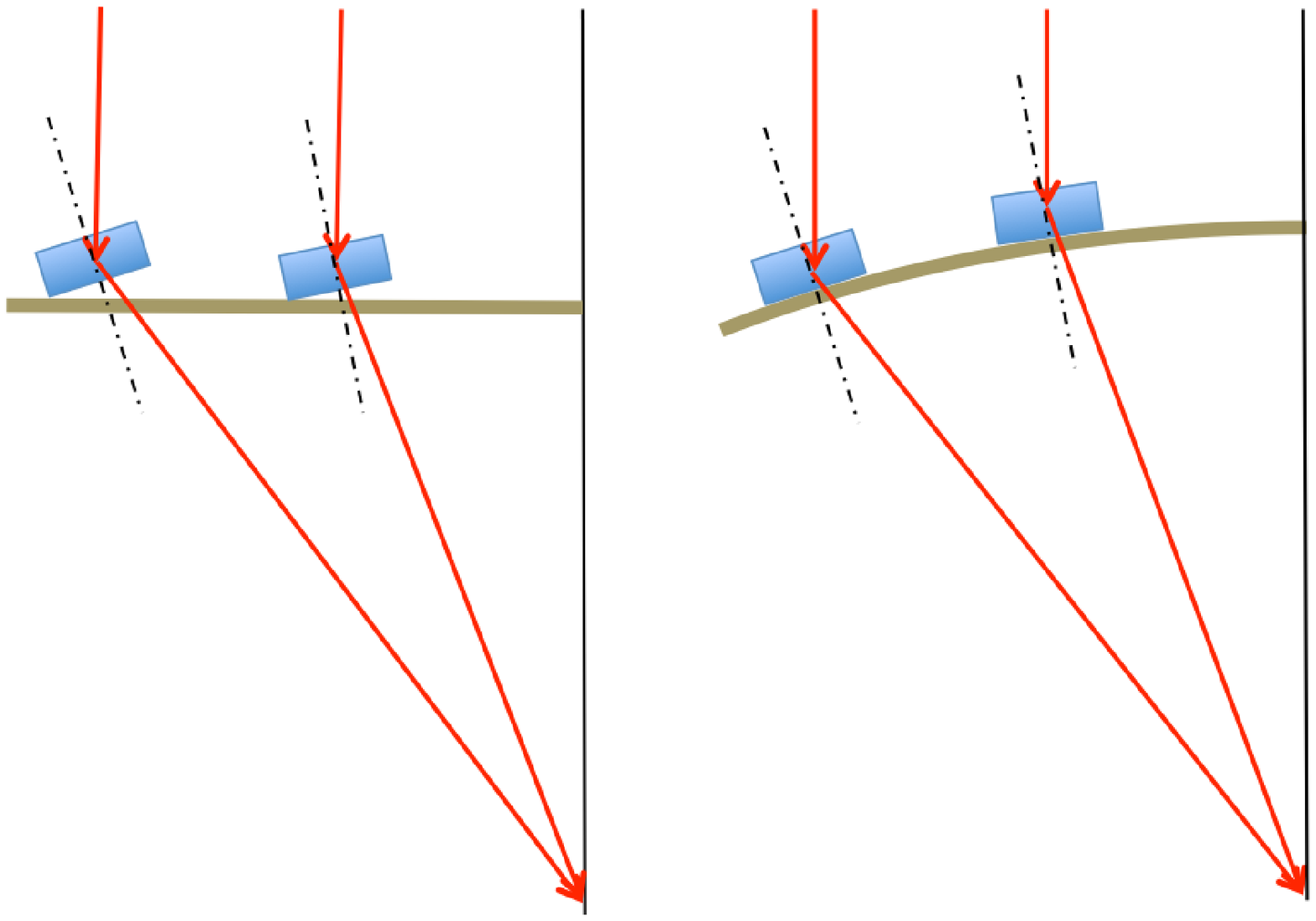}
\includegraphics[width=0.5\textwidth]{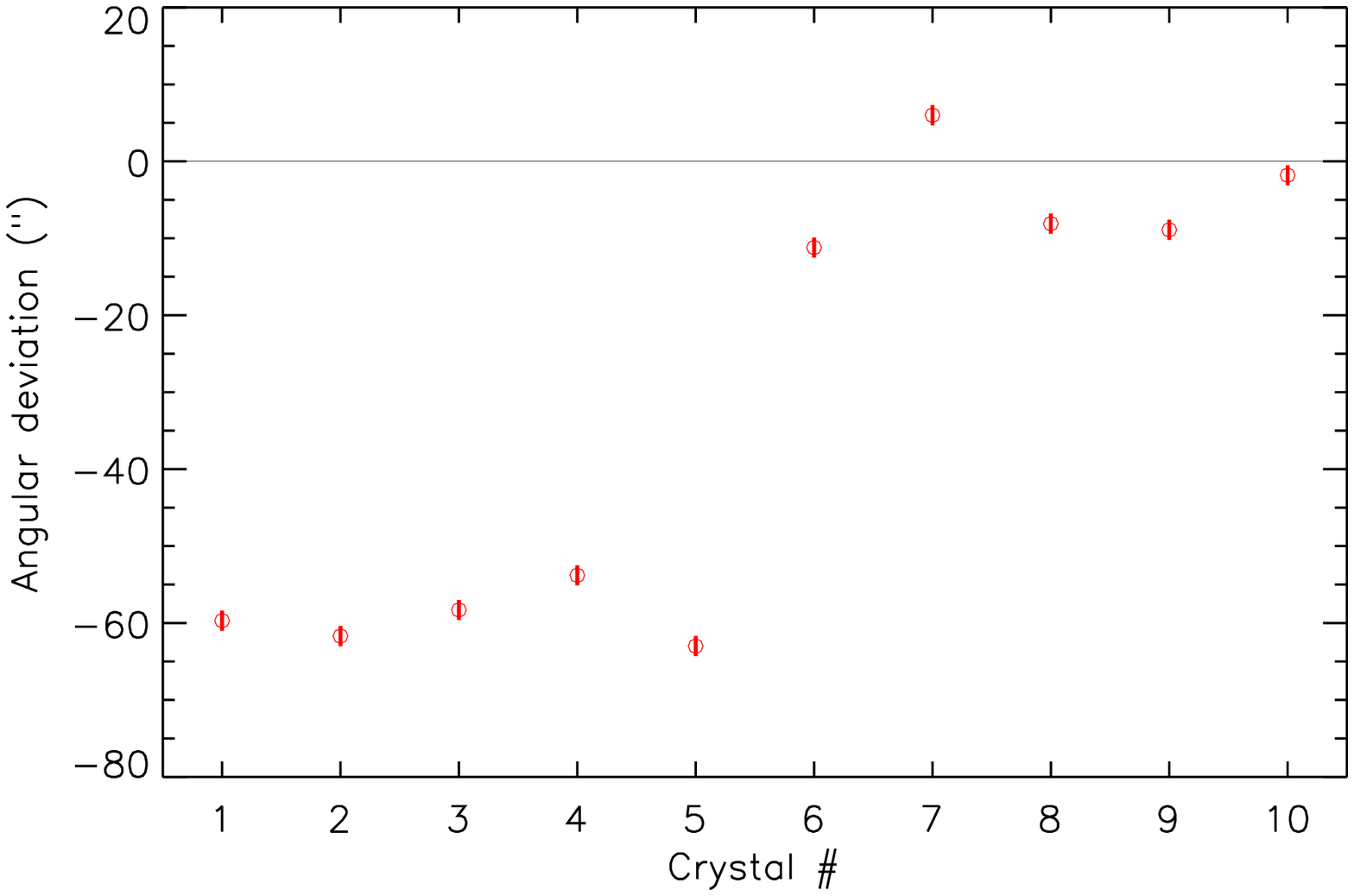}
\caption{{\it Left:} Comparison between the two Laue lens substrate options tested. On the left, the substrate is a flat disc while on the right it is a paraboloid.
{\it Right:} Angular deviation with respect to the desired orientation for each of the 10 crystals glued on the aluminum substrate.}
\label{fig:results}
\end{center}
\end{figure}

The crystals are arranged in two partial concentric rings: For the inner ring the substrate was set roughly perpendicular to the beam, which means that there is a wedge of glue of about 1.13$^{\circ}$ behind each crystal due to the Bragg angle. For the second ring, however, the substrate was brought nearly parallel to the oriented crystals, making the bond line nearly parallel. The aim was to test two different substrate geometries, either flat (and the Bragg angle is created by a wedge of glue) or paraboloid, as first proposed by Lindquist [\citenum{lindquist.1968az}], where the normal to the substrate stays parallel to the diffracting planes of crystals at any radius (Figure \ref{fig:results}). In the case of crystals cleaving along their crystalline planes, the latter case would in theory allow for just laying the crystals on the substrate and having them be correctly oriented. Unfortunately, we do not know any crystal that is efficient for diffraction at high energy that also cleaves. 

In the flat substrate case, we noticed that the glue shrinkage upon cure induced a very reproducible deviation of $-$60 $\pm$ 3 arcsec. This is surprising as a Bragg angle of 1.133$^{\circ}$ implies for a 5-mm large crystal a wedge of $\sim$100 $\mu$m. According to the glue manufacturer, the glue shrinkage should then be of 0.03 $\mu$m, which corresponds to an anglular deviation of 1.2 arcsec. We need to investigate why we observed a deviation 50 times larger.
However, the very positive result is the 3 arcsec standard deviation over these 5 crystals. This shows that our method produces reproducible results.

In the 'paraboloid' substrate case, the average deviation over the five crystals is $-$5 $\pm$ 6 arcsec. The residual offset is attributed to a small glue wedge (the substrate was roughly turned to be parallel to an oriented crystal). Again the good result is the small variation between the five crystals: only 6 arcsec.

It is to be noticed that this trial was conducted in the high bay of SSL, with the Laue lens assembly station being 1 m away of the gate that is exposed to the sun in the afternoon. We observed temperature variations in the range 22.0$^{\circ}$C - 29.5$^{\circ}$C during the gluings. Although the influence of the temperature variation was not clearly established, we believe a light thermal control would increase the stability of the setup.

Three crystals over the course of 24 hours could be glued with the present method and glue. Setting up a crystal at the tip of the vacuum chuck, getting a diffraction spectrum (which tells us if this crystal is good enough as well as its orientation), then bringing it in gluing position and injecting the glue takes less than 30 min. With a glue curing in one hour, a dozen crystals per day could be mounted. Now that we demonstrated that it is possible to fix crystals with accuracy better than 10 arcsec and dense packing factor, our next endeavor will be to speed up the process.

 \section{CONCLUSIONS}
 
 At SSL, we are taking up the challenge of the development of the first scientifically viable gamma-ray optic, based on densely packed, accurately oriented, efficient crystals. The very first test of our Laue lens assembly setup and process showed excellent results. Soon, a second trial will take place with a refined substrate (paraboloid). The glue we have been using takes six hours (at least) to cure. We estimate that it is necessary to decrease this time down to at the most two hours in order to be able to assemble an astronomical Laue lens within one to two years. Indeed, in most current projects, the lens is composed of at least 5000 crystals. 
Now that we have proven that the required precision can be met, we will work to speed up the process. 

In parallel we are conducting studies with Mateck to enable high efficiency crystals for high energies. We hope that this activity will demonstrate the availability of pure rhodium, silver and lead crystals and consequently raise their Technology Readiness Level.  The 50 crystals we ordered will be delivered in Spring 2012.

Laue lenses telescopes are not designed to be general purpose observatories, but rather to answer very specific questions requiring one to two order of magnitude sensitivity improvement with respect to existing instruments. Black holes (both stellar mass and super-massive), Galactic positron origin and Type Ia supernovae are three examples of topics that would greatly benefit from the advent of Laue lens telescopes.

\acknowledgments     
 
The authors wish to thank Dr. Nikolay Abrosimov (IKZ, Germany) for providing the Si crystals that were used in this study. The authors also wish to thank Dr. Michael Jentschel, Gilles Roudil and Dr. Pierre Bastie for their precious help during the characterization of the silver, rhodium and lead ingots at ILL on the DIGRA beamline.



\begin{thebibliography}{10}

\bibitem{lund.1992ft}		
N.~Lund.
``{A study of focusing telescopes for soft gamma rays}'', {\em Exp. Astronomy}, {\bf2}:259--273 (1992).

\bibitem{barriere.2010fk}
N.~{Barri{\`e}re}, L.~Natalucci and P.~Ubertini.
 ''{Hard X-ray/soft gamma-ray polarimetry using a Laue lens}'', 
 in ''{X-ray Polarimetry: A New Window in Astrophysics}''
{\em Cambridge university press edition}, p. 88 (2010).

\bibitem{frontera.2005uq}
F.~{Frontera}, A.~{Pisa}, P.~{de Chiara}, and {et al.}
``{Exploring the hard X-/soft gamma-ray continuum spectra with Laue lenses}''.
 {\em 39TH ESLAB Symposium on Trends in Space Science and Cosmic
  Vision 2020}, vol. {\bf 588} of {\em ESA Special Publication}, p. 323 (2005).

\bibitem{barriere.2009fk}
N.~M. {Barri{\`e}re}, L.~{Natalucci}, N.~{Abrosimov}, P.~{von Ballmoos},
  P.~{Bastie}, P.~{Courtois}, M.~{Jentschel}, J.~{Kn{\"o}dlseder},
  J.~{Rousselle}, and P.~{Ubertini}, III.
``{Soft gamma-ray optics: new Laue lens design and performance estimates}''.
  {\em Proc. SPIE}, {\bf7437}: 74370K--74370K--11 (2009).

\bibitem{barriere.2008}
N.~{Barri{\`e}re}, L.~{Natalucci}, and P.~{Ubertini}.
``{Soft gamma-ray polarimetry using a Laue lens telescope: The gamma-ray
  imager}'', {\em Proceedings of Science}, PoS(CRAB2008)019 (2008).

\bibitem{von-ballmoos.2010zr}
P.~{von Ballmoos}, T.~{Takahashi}, and S.~E. {Boggs}.
``{A DUAL mission for nuclear astrophysics}'', {\em Nuclear Instruments and Methods in Physics Research A},
  {\bf 623}:431--433 (2010).

\bibitem{boggs.2010uq}
S.~{Boggs}, C.~{Wunderer}, P.~{von Ballmoos}, T.~{Takahashi}, N.~{Gehrels},
  J.~{Tueller}, M.~{Baring}, J.~{Beacom}, R.~{Diehl}, J.~{Greiner}, E.~{Grove},
  D.~{Hartmann}, M.~{Hernanz}, P.~{Jean}, N.~{Johnson}, G.~{Kanbach},
  M.~{Kippen}, J.~{Kn{\"o}dlseder}, M.~{Leising}, G.~{Madejski},
  M.~{McConnell}, P.~{Milne}, K.~{Motohide}, K.~{Nakazawa}, U.~{Oberlack},
  B.~{Phlips}, J.~{Ryan}, G.~{Skinner}, S.~{Starrfield}, H.~{Tajima},
  E.~{Wulf}, A.~{Zoglauer}, and A.~{Zych}.
``{DUAL Gamma-Ray Mission}'',
{\em astro2010: The Astronomy and Astrophysics Decadal Survey} (2010).

\bibitem{barriere.2010rt}
N.~M. {Barri{\`e}re}, J.~A. {Tomsick}, and S.~E. {Boggs}. 
``{Unveiling physical processes in type ia supernovae with a Laue lens telescope.}'' 
{\em PoS(INTEGRAL 2010)108} (2010).

\bibitem{knodlseder.2009kx}
J.~{Kn{\"o}dlseder}, P.~{von Ballmoos}, F.~{Frontera}, A.~{Bazzano},
  F.~{Christensen}, M.~{Hernanz}, and C.~{Wunderer}. 
  ``{GRI: focusing on the evolving violent universe}'', {\em Exp. Astronomy} {\bf 23}(1):121--138 (2009).

\bibitem{leising.2005qr}
M.~D. {Leising}. 
``{Focusing supernova gamma rays.}'', {\em Exp. Astronomy} {\bf20}:49--55 (2005).

\bibitem{knodlseder.2005rc}
J.~{Kn{\"o}dlseder}, P.~{Jean}, V.~{Lonjou}, G.~{Weidenspointner},
  N.~{Guessoum}, W.~{Gillard}, G.~{Skinner}, P.~{von Ballmoos}, G.~{Vedrenne},
  J.-P. {Roques}, S.~{Schanne}, B.~{Teegarden}, V.~{Sch{\"o}nfelder}, and
  C.~{Winkler}. 
``{The all-sky distribution of 511 keV electron-positron annihilation emission}.'' {\em Astronomy \& Astrophysics} {\bf 441}:513--532 (2005).

\bibitem{abdo.2010fk}
A.~A. {Abdo}, {\it et al.}.
``{Fermi Large Area Telescope View of the Core of the Radio Galaxy
  Centaurus A}.''
{\em ApJ} {\bf 719}:1433-1444 (2010).

\bibitem{grove.1998kx}
J.~E. {Grove}, W.~N. {Johnson}, R.~A. {Kroeger}, K.~{McNaron-Brown}, J.~G.
  {Skibo}, and B.~F. {Phlips}.
``{Gamma-Ray Spectral States of Galactic Black Hole Candidates}''.
{\em ApJ} {\bf 500}:899-+ (1998).

\bibitem{bouchet.2009uq}
L.~{Bouchet}, M.~{del Santo}, E.~{Jourdain}, J.~P. {Roques}, A.~{Bazzano}, and
  G.~{DeCesare}.
``{Unveiling the High Energy Tail of 1E 1740.7-2942 With INTEGRAL}''.
{\em ApJ} {\bf 693}:1871-1876 (2009).


\bibitem{coburn.2001fk}
W.~{Coburn}, S.~E. {Boggs}, S.~{Amrose}, R.~P. {Lin}, M.~T. {Burks},
  M.~{Amman}, P.~N. {Luke}, N.~W. {Madden}, and E.~L. {Hull}.
``{Results of charge sharing tests in a ge-strip detector.}'', {\em IEEE Transactions on Nuclear Sciences} {\bf 1}:226--229 (2001).


\bibitem{barriere_2010b}
N.~M. {Barri{\`e}re}, J.~{Tomsick}, S.~E. {Boggs}, J.~{Rousselle}, and P.~{von
  Ballmoos}. 
``{Development of efficient Laue lenses: experimental results and projects.}'', {\em Proc. SPIE} {\bf 7732}:773226--773226--11 (2010).

\bibitem{bellm.2009kx}
E.~C. {Bellm}, S.~E. {Boggs}, M.~S. {Bandstra}, J.~D. {Bowen},
  D.~{Perez-Becker}, C.~B. {Wunderer}, A.~{Zoglauer}, M.~{Amman}, P.~N. {Luke},
  H.-K. {Chang}, J.-L. {Chiu}, J.-S. {Liang}, Y.-H. {Chang}, Z.-K. {Liu}, W.-C.
  {Hung}, C.-H. {Lin}, M.~A. {Huang}, and P.~{Jean}. 
``{Overview of the Nuclear Compton Telescope}'', {\em IEEE Transactions on Nuclear Sciences} {\bf 56}:1250--1256 (2009).

\bibitem{von-ballmoos.2004sf}
P.~{von Ballmoos}, H.~{Halloin}, J.~{Evrard}, G.~{Skinner}, N.~{Abrosimov},
  J.~{Alvarez}, P.~{Bastie}, B.~{Hamelin}, M.~{Hernanz}, P.~{Jean},
  J.~{Kn{\"o}dlseder}, V.~{Lonjou}, B.~{Smither}, and G.~{Vedrenne}.
``{CLAIRE's first light}'', {\em New Astronomy Review} {\bf 48}:243--249 (2004).


\bibitem{Barriere:he5432}
N.~M. Barri{\`{e}}re, J.~Rousselle, P.~von Ballmoos, N.~V. Abrosimov, P.~Courtois,
  P.~Bastie, T.~Camus, M.~Jentschel, V.~N. Kurlov, L.~Natalucci, G.~Roudil,
  N.~Frisch~Brejnholt, and D.~Serre. 
``{Experimental and theoretical study of the diffraction properties of
  various crystals for the realization of a soft gamma-ray Laue lens}'', {\em Journal of Applied Crystallography} {\bf 42}(5):834--845  (2009).

\bibitem{rousselle.2011fk}
J.~Rousselle, P.~von Ballmoos, M.~Jentschel, N.~V. Abrosimov, N.~M. Barri{\`{e}}re,  P.~Courtois, and G.~Roudil.
``{Cu and SiGe gradient crystal productions and measurements for a Laue lens application}'', {\em submitted to Experimental Astronomy}.


\bibitem{lindquist.1968az}
T.~R. Lindquist and W.~R. Webber.
``{A focusing x-ray telescope for use in the study of extraterrestrial x-ray sources in the energy range 20-140 kev.}'', {\em Canadian Journal of Physics} {\bf 46}:1103 (1968).


\end{thebibliography}
\bibliographystyle{spiebib}   

\end{document}